\documentclass[apjl]{emulateapj}

\usepackage{apjfonts}

\usepackage{amsmath}
\usepackage{natbib}

\usepackage{epsfig}

\newcommand{\yr}{\ensuremath{{\rm yr}}} 
\newcommand{\cm}{\ensuremath{{\rm cm}}} 
\newcommand{\kpc}{\ensuremath{{\rm kpc}}}

\newcommand{\GeV}{\ensuremath{{\rm GeV}}} 
\newcommand{\TeV}{\ensuremath{{\rm TeV}}} 
\newcommand{\erg}{\ensuremath{{\rm erg}}} 

\newcommand{\MHz}{\ensuremath{\, {\rm MHz}}}
\newcommand{\GHz}{\ensuremath{\, {\rm GHz}}}
\newcommand{\muGs}{\ensuremath{\, \mu{\rm Gs}}}

\newcommand{\etal}{{\it et al.}}

\begin{document}

\title{Angular power spectrum of the galactic synchrotron radiation}

\author{Xuelei Chen}
\affil{The Kavli Institute for Theoretical Physics, UCSB, Santa
Barbara, CA 93106, U.S.A.}

\begin{abstract}
We calculate the angular power spectrum of the galactic synchrotron
radiation induced by the small scale fluctuations of the magnetic field
and the cosmic ray electron density. 
Using the observed interstellar magnetic field 
spectrum, which is consistent with the Komolgorov turbulence model
at the relevant scales, we find that $C_l \propto
l^{-3.7}$. We estimate the cosmic ray 
electron density fluctuation spectrum with
an injection-diffusion model, the shape of the angular power spectrum 
in this model depends on the correlations between the injection
sources. For Poisson distribution of sources, $C_l \propto l^{-4}$.
We discuss the implications for the interpretation of cosmic
microwave background (CMB) data and the impact on future 21 cm
tomography experiments. 
\end{abstract}

\keywords{ cosmic microwave background --- 
radiation mechanism:  non-thermal --- radio continuum:  galaxies
---ISM: magnetic fields --- cosmic rays}

\section{Introduction}

The radio sky at $\nu < 1 \GHz$ is dominated by galactic synchrotron
emission. It is believed to be produced by cosmic ray electrons propagating
in the magnetic field of the Galaxy \citep{GS69}.
The galactic synchrotron emission is an important foreground 
for the cosmic microwave background (CMB) 
experiments \citep{S99}. For the upcoming high redshift 
21cm tomography experiments \citep*{MMR97,TMMR00,CM04,CM03,GS03,
F04a,ISMS03,LZ04}, such as 
PAST \citep{PWP04}, LOFAR\footnote{http://www.lofar.org}, and 
SKA\footnote{http://www.skatelescope.org}, it poses a major challenge.

The global galactic synchrotron emission spectrum from 0.3 MHz to 408
MHz can be fit with a two component disk model of 
the galaxy \citep*{KWL04a,KWL04b}.  
Analysis of the existing radio surveys at 408 MHz, 1.42 GHz, and 2.326
GHz \citep*{H82,R82,RR86,RR88,JBN98} shows that the synchrotron emission 
has a spectral index $\beta \approx 2.7$ \citep{P03}, which is in general
agreement with the CMB result \citep{B03}. The real space 
distribution of synchrotron emissivity over the galactic disk, 
taking into account spiral arms, were derived from the 408 MHz whole sky
map using refolding techniques \citep*{P81a,P81b,BKB85}.
In the Fourier space, the angular power spectrum of the galactic 
synchrotron radiation can be modeled reasonably well 
as a simple scaling relation: 
$C_l \propto l^{-\eta} $ \citep*{TE96,TEHO00,G01,G02}, with $\eta = 2.4-3$
down to $l \sim 900$. Recently, the WMAP team obtained a shallower
spectrum of $\eta \sim 2$ down to $l \sim 200$. These are extrapolated
to higher $l$ in recent studies of the 21 cm foreground \citep*{DMCM04,SCK04}.

In the present study we investigate the galactic synchrotron emission from a 
different perspective. We calculate the synchrotron 
angular power spectrum induced by the variation of the magnetic field
and fluctuations in the cosmic ray electron density.  
Our primary objective is to achieve a {\it physical} understanding of 
the origin of the synchrotron emission anisotropy. In
particular, we would like to ask which of these two mechanisms 
is responsible in producing the observed anisotropy? 
This will help us to assess the validity of the hypothesis 
adopted in the empirical analyses of CMB and 21 cm observation, e.g., 
will the power law form of the angular power spectrum hold
down to small scales relevant for the pre-reionization 21 cm observation?
At the same time, we may also gain useful knowledge about the galactic
distribution of the cosmic ray electrons and the magnetic field.
Similar {\it  physical} modeling have been performed for a number of 
other foregrounds, e.g. free-free \citep{OM03,CF04} and intergalactic
shocks \citep{KWL04a,KWL04b}. 

The fluctuation power spectrum of the galactic magnetic field has been 
measured \citep*{MS96,hfm04}, and is consistent with being produced by 
a turbulent interstellar medium (ISM) described by the 
Komolgorov scaling model \citep{K41,ES04}. 
With this we can calculate directly the anisotropy power spectrum 
induced by the magnetic field variation. The distribution of the 
cosmic ray electrons is less well known, but 
there is a broadly accepted picture of the cosmic electrons being produced in 
supernovae remnants (SNR), which then diffuses through the whole galaxy and 
be confined in a volume greater than the galactic disk \citep{GS64}.
We can calculate the cosmic ray density distribution with this model.

\section{Models}

Let us consider the angular power spectrum obtained over a small patch of
``blank'' sky field at high galactic latitude.
In the optically thin case, the observed radiation intensity 
is simply an integral of the 
emissivity along the line of sight. At very low frequencies, synchrotron
self-absorption and plasma absorption are important. For $70 \MHz< \nu <
200 \MHz$, corresponding to the 
redshift range of 6-20 which is of interest to the study of the 
reionization process, these can be neglected,
\begin{equation}
I(\hat{n},\nu) = \int dr ~ \varepsilon(\hat{n}, r, \nu), 
\end{equation}
where $\epsilon$ is the volume emissivity. 
The emissivity may have both spatial and frequency variations. If the two
are separable, we may write 
\begin{equation}
\varepsilon_{\nu}({\bf x})= g(\nu) \psi({\bf x}), \qquad 
\tilde\varepsilon_{\nu}({\bf k})=g(\nu) \tilde \psi({\bf k});
\end{equation}
 where the tilde denotes Fourier transform, 
$\psi({\bf x}) = \int \frac{d^3 k}{(2\pi)^3} e^{i{\bf k}\cdot {\bf x}} 
\tilde{\psi}({\bf k})$.
The emissivity  power spectrum is then 
\begin{equation}
P_{\varepsilon}(\nu_1, \nu_2, \mathbf{k}) = g(\nu_1) g(\nu_2) P_{\psi}(\mathbf{k}),
\end{equation}
Let us consider the case $\nu_1=\nu_2=\nu$. 
At small scale, using the Limber
approximation \citep{K92,ZFH04}, the angular power spectrum is given by
\begin{equation}
\label{eq:cl}
C_l^T (\nu)= \frac{c^4 g^2(\nu)}{4\nu^4 k_B^2}
  \int \frac{dr}{r^2} P_{\psi}(\frac{l}{r}).
\end{equation}
If the emissivity has a power law spectrum, $P_{\psi}(k) \propto k^{\eta}$,
then $C_l \propto l^{\eta}$. The integral is up to some cutoff point,
at which the emissivity drops to 0.

We now consider the emissivity of synchrotron radiation. 
If the energy distribution of cosmic ray
electrons $f(E)$ at each point is approxiamated as a simple power law with 
$f(E) = C E^{-p}$,
then 
\begin{equation}
\varepsilon(\nu) =\frac{\sqrt{3} e^3}{m_e c^2} C B_{\perp} \alpha_{syn}(p)
\left(\frac{2\pi m_e c\nu}{3eB_{\perp}}\right)^{-(p-1)/2}
\end{equation}
where $\alpha_{syn}(p)=\frac{1}{p+1}\Gamma[\frac{3p-1}{12}] 
\Gamma[\frac{3p+19}{12}]$.
We see that for power law distribution with $f(E) \propto E^{-p}$, 
$I(\nu) \propto \nu^{\alpha}$, and $T(\nu) \propto \nu^{\beta}$, where 
$\alpha=-(p-1)/2$, and $\beta=\alpha-2$. A
power law is actually a good approximation to the real
distribution function of cosmic ray electrons. 
In this approximation, variation of emissivity can be
induced by varying the magnetic field $B$, spectral index $p$, and
cosmic ray electron density normalization $C$. 
Here we shall fix $p$ and consider the variation of $B$ and $C$.

The galactic magnetic field is typically a few $\muGs$,  for this
magnetic field the emission at $70-200 \MHz$ is produced primarily 
by cosmic electrons with $E \sim 0.1 ~\GeV - 10 ~\GeV$. 
The local cosmic ray electron density 
can be measured directly. There is some uncertainty in the
normalization, and corrections have to be made for solar modulation.  
A recent compilation of measurements yields \citep{CB04}, in our units,  
$C=1.7 \times 10^{-11} \cm^{-3} \GeV^{-1}$, 
and $p=3.44$, at the energy of a few GeV.
The spectral index $p=3.4$ yields $\alpha= -1.2$, and $\beta=
-3.2$. For comparison, the WMAP measurement \citep{B03} 
indicates that towards star 
forming regions, $\beta=-2.5$, and towards the halo,
$\beta=-3.0$. Radiative loss of electron energy provides a natural
explanation to the steeping of the electron spectrum away from
star-forming region. Below 3 GeV, the electron spectrum becomes
flatter, probably because the primary energy loss mechanism changes from 
radiation to ionization.
The locally measured electron density may not 
represent the average, nevertheless we will use it as a trial value.
We will take $p=3$, which gives a slightly better fit to the radio
data than the steep $p=3.4$ value. 
We also assume a disk scale height of 1 kpc (corresponding to the
thick disk in the \citealt{KWL04a} model), and a smooth magnetic field
of $4 \muGs$. With this set of parameters, the integrated sky
brightness temperature at 408 MHz is 20 K, which reproduces the
observed value at high galactic latitude \citep{H82}.

Let us consider first the magnetic field with $C$ fixed. Now 
\begin{equation}
\psi= b(x) = B_{\perp}^{(p+1)/2}(x).
\end{equation}
If there is a large smooth global magnetic field component which
varies only on large scale and 
fluctuations around it are small, then we can write $B=B_0+\delta
B(x)$, and 
\begin{equation}
B_{\perp}^{(p+1)/2}=B_{0\perp}^{(p+1)/2}\left(1+\frac{p+1}{2}
\frac{\delta B_{\perp}(x)}{B_{0\perp}} + ...\right)
\end{equation}
then
\begin{equation}
P_{\psi}(k) = \frac{(p+1)^2}{6} C^2 B_{0\perp}^{p-1} P_{\delta B} (k) .
\end{equation}

\begin{figure}
\plotone{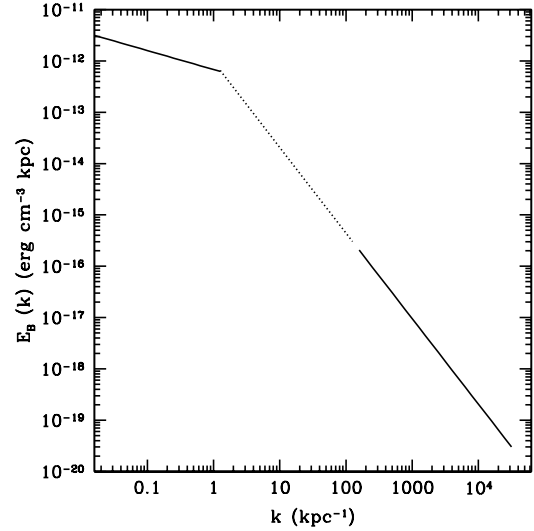}
\caption{\label{fig:Bspec} The magnetic field energy spectrum as extrapolated
  from \cite{MS96} and \cite{hfm04}. The dotted part is interpolation.}
\end{figure}

The interstellar medium is turbulent \citep{ES04}. Komolgorov derived
a scaling relation for scale-invariant turbulence \citep{K41}, 
with a power spectrum of the form $k^{-11/3}$.
The magnetic field fluctuation spectrum can be determined by
energy equipartition. The predicted magnetic field fluctuation 
spectrum is confirmed on the 
scales of 0.01 pc -- 100 pc by observation of the Faraday rotation 
of extragalactic sources \citep{MS96}. At larger scales and on the
galactic disk, the magnetic field spectrum is flatter, probably
because at these scales the motion is dominated by 
by two-dimensional structure (vortices). If we join the 
small scale and the large scale observations \citep{hfm04}, we obtain
\begin{equation}
\label{eq:Bspec}
E(k) = \left\{ \begin{array}{ll}
2.03 \times 10^{-11} ~
k_{\kpc^{-1}}^{-5/3}
~\erg~ \cm^{-3}~ \kpc,& k_{\kpc^{-1}} 
> 1.57 \times 10^3 \\
1.34 \times 10^{-12} ~
k_{\kpc^{-1}}^{-0.37}
\erg~ \cm^{-3}~ \kpc, & k_{\kpc^{-1}}  < 6.28 \\
\end{array}\right.
\end{equation}
This is plotted in Fig.~\ref{fig:Bspec} (Note that our definition of k differs from \cite{hfm04} 
by a factor of $2\pi$). The energy is related to the
power spectrum by 
\begin{equation}
P_B(k) = 2 E_B(k)/k^2.
\end{equation} 
Other scaling models have also been suggested. For example, some MHD
turbulence have $E \sim k^{-3/2}$ instead of $k^{-5/3}$ \citep{K65}.
However, this would produce an angular power spectrum not very
different from the Komolgorov one.

We can then carry out the calculation, with 
the magnetic field spectrum given in Eq.~(\ref{eq:Bspec}) and 
the locally measured electron density. However, 
even though we keep $C$ fixed to investigate the variation induced by
the magnetic field, in reality it must decrease as we move away from
the galactic disk. We can approximate this effect by taking the power
spectrum $P_{\psi}(k)$ also as an explicit function of $r$, i.e. 
$P_{\psi}(k,r) \sim P(k) e^{-2r/r_0}$, where
$r_0$ is the scale height of the halo in which the cosmic ray
electrons are confined. It turns out that this damping factor only
changes the result by a small factor, because the large $r$
contribution is already suppressed by the $P \sim (l/r)^{-5/3}$ factor.
For the small scale that we are mostly interested in, Komolgorov
spectrum applies, $P_B(k) \sim k^{-11/3}$, and
$C_l \sim l^{-3.7}$. 

\begin{figure}
\plotone{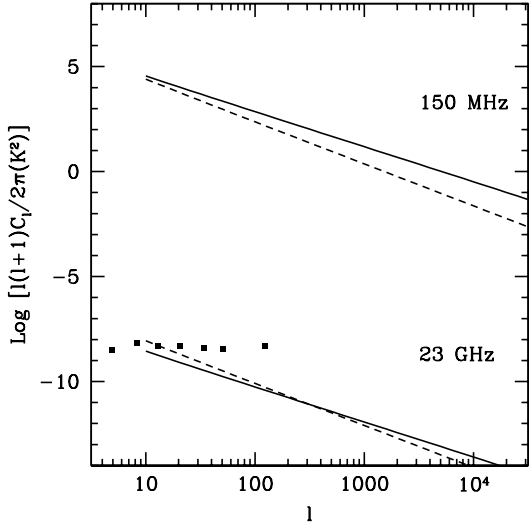}
\caption{\label{fig:cl} The synchrotron power spectrum for
  $\nu=150\MHz$ (upper curves) and   $\nu=20\GHz$ (lower curves). 
The solid line marks the anisotropy induced by the 
fluctuation of magnetic field, the dashed line for that induced by the
fluctuation of cosmic ray
electron density, the points are WMAP K band (23 GHz) data. } 
\end{figure}

The result of our calculation for 
$\nu=150 \MHz$ (21 cm line at $z \sim 9$), 
and $ 23 \GHz$ are plotted in Fig.~\ref{fig:cl} as solid lines, along
with the WMAP K band data (centered at 23 GHz). 
Remarkably, at small $l$, the anisotropy power amplitude is 
of the same order of magnitude as the WMAP data (At such small $l$,
the Limber approximation may not be very good, nevertheless the true 
result will be of the same order of magnitude). This means that the 
magnetic field fluctuation may play a role in the formation of the
observed synchrotron anisotropy. However, it is clear that compared 
with the data $C_l$ drops too fast as $l$ increases.

If the smooth magnetic field component is absent and the 
fluctuating field dominates, then 
\begin{equation}
P_{\psi} = \langle B_{\perp}^{p+1} \rangle (k) = 
\langle B^{p+1} \rangle (k) c(p) 
\end{equation}
where 
$c(p)= \frac{1}{2}\int_0^{\pi} d\theta ~ \sin^{p+2}\theta=
\sqrt{\pi}~ \Gamma[(3+p)/2]/\Gamma[(2+p/2)]$.
In this case, the result is uncertain, 
because it depends on the higher order
correlation $\langle P_{B^{(p+1)/2}}(k) \rangle $. 
If we make the gaussian-like {\it ansatz} $\langle B^{p+1}(k)\rangle 
\sim (\langle B^2(k)\rangle)^{(p+1)/2} $, and $P_B(k) \sim k^{-\eta_B}$,
then the result is 
$C_l \propto l^{-(p+1)\eta_B/2}$. For
the Komolgorov spectrum $\eta_B = -11/3$, $C_l \sim l^{-7.3}$, 
which is extremely steep and can be neglected entirely. However, this 
assumption maybe incorrect.

Now we consider the variation of the cosmic ray electron density. This
is not known from observation. To make an estimate of the fluctuation, 
we consider an injection-diffusion model \citep{GS64}, in which 
the cosmic ray electrons are produced (injected) in point sources,
then diffuse out, until eventually losing all their energy or 
escaping the confinement volume. The density of electrons at a point in
space is then dependent on the distance to nearby sources. 
Neglecting the momentum space diffusion, the density
normalization constant at position $\mathbf{r}$, time $t$ satisfies
the diffusion equation
\begin{equation}
\frac{\partial C(\mathbf{r},t)}{\partial t} = D \nabla^2
C(\mathbf{r},t) + q(\mathbf{r},t) - \frac{C(\mathbf{r},t)}{\tau}
\end{equation} 
where $q(\mathbf{r},t)$ is the cosmic ray electron injection rate at
$\mathbf{r}$, and $D$ is the diffusion coefficient, 
which is constant below 5 GeV,
and $D \approx 2\times 10^{28} (E/5\GeV)^{0.6}$ at $E > 5 \GeV$
\citep{KKYN04}. The last term in the equation represents loss of
electrons by radiation, ionization, etc., or by escaping the
confinement volume, with $\tau$ the loss time scale. For radiative
losses \citep{CB04}, 
\begin{equation}
\tau \approx 2.1 \times 10^5 (E/\TeV)^{-1} \yr.
\end{equation}
We note that in the injection-diffusion model, the scale height of the 
cosmic ray halo is a few times of 
$\times \sqrt{D \tau}$,  with the above values we have 
$\sqrt{D\tau} \sim 0.3 \kpc$. 
Although very crude, our model is self-consistent. 
In Fourier space, 
\begin{equation}
\frac{\partial \tilde{C}(k)}{\partial t} + (D k^2+\frac{1}{\tau})
\tilde{C}(k) = \tilde{q}(k)
\end{equation}
The steady state solution is 
\begin{equation}
\tilde{C}(k) = \frac{\tilde{q}(k)}{D k^2 +\frac{1}{\tau}}.
\end{equation}
The power spectrum is then
\begin{equation}
P_C(k) = \frac{P_q(k)}{(D k^2 +\frac{1}{\tau})^2}.
\end{equation}

Supernovae remnants are most likely the primary source for these cosmic
ray electrons. The injection function is then $q = N_e \kappa_{SN}$, 
where $N_e$  is the number of 
cosmic ray electrons produced in one supernova, 
and  $\kappa_{SN}$ is the number of supernovae explosions per unit
volume per unit time. If the distribution of supernovae is Poisson, 
with $\langle \kappa_{SN} \rangle =1/(V t_{SN})$, 
where $V$ is the volume in which 
one supernova explode per average interval $t_{SN}$, then
\begin{equation}
P_C(k) = \frac{N_e^2 V}{(D k^2 +\frac{1}{\tau})^2 t_{SN}^2 V^2} 
\end{equation}

When $k \to 0$, we have $P_C(k) = C_0^2 V$, where $C_0$ is the average
of $C$. From this we obtain 
\begin{equation}
\frac{N_e}{V} = C_0 \frac{t_{SN}}{\tau}
\end{equation}
As a reality check, we take $E \sim 1 \GeV$, 
a stellar disk with radius of 15 kpc,
and a scale height of 300 pc, and also assume that in this volume the 
average interval of supernovae explosion is 50 years, then we find 
$N_e \sim 10^{48} $, which requires an energy of $10^{45} \erg$, 
which is a small fraction of the total energy of a supernova, hence
supernovae do have sufficient energy to generate these cosmic ray electrons.
 
Using this relation we finally obtain
\begin{equation}
P_C = \frac{C_0^2}{ (D k^2 \tau +1)^2} V,
\end{equation}
Note that when written in this form, 
the result does not depend on $t_{SN}$.
The resulting $C_l$ is plotted as dashed curves in Fig.~\ref{fig:cl}.
At large angle, the predicted $C_l$ in this model is comparable to the
case of magnetic field induced fluctuation, and agrees with what is
observed at the order of the magnitude level. However, at large $l$,
$C_l \sim l^{-4}$,  again the spectrum is too steep compared with
observation. Although the spectrum flattens at small $k$, this
happened only at scales comparable to the disk scale height, thus
affecting only large angle ($l \sim$ a few). 

In the above we have assumed a Poisson distribution of supernovae
remnant. If they are correlated, with
$P_{SN}(k) \propto A k^{\gamma}$, then
$C_l \propto  l^{\gamma-4} $. 
Supernovae may well be correlated, as their rate should be 
proportional to the star formation rate, which in turn depends on the 
density. However, the Komolgorov model of turbulence suggests
$\gamma<0$. The observation of flocculent star light distribution 
in nearby spiral galaxies seem to confirm this expectation,
which has a power law of $P_{k} \sim k^{-1}$ at the 100 pc 
scale \citep*{EEL03,E03}. If so, then the correlation of SNR may not
help us. However, more observations are needed to address this
issue.

\section{Conclusion}
We have calculated angular power spectrum of the galactic synchrotron
radiation induced by the variation of the magnetic field and cosmic
ray electron density. We found that at low $l$, the amplitudes of the 
anisotropy power produced by both mechanisms are comparable to the
observed value. This indicates that these physical mechanisms 
{\it are} relevant to the production of the observed foreground anisotropy.  
However, neither of these two mechanisms can produce the 
$C_l \sim l^{-2}$ power spectrum observed by WMAP \citep{B03}, and are also
steeper than the older value of $\eta \sim 2.4-3 $. 
The magnetic field model induce power spectrum of the form 
$C_l \sim l^{-3.7}$, while for the electron density fluctuation 
it is $C_l \sim l^{-4}$ if the spatial distribution of the SNR is
Poisson. This may be remedied if the spatial distribution of supernova
remnants has the form $P_{SN}(k) \sim k^2$. However, observations seem
to indicate $P_{SN}(k) \sim k^{-1}$.

We made a number of simplifications in our calculation. 
We did not consider the detailed distribution of global
magnetic field and cosmic ray sources in the Galaxy, nor do we
consider accelerations outside SNR. It is unlikely that inclusion of 
any of these details would change our qualitative conclusion.  
Perhaps more important is our assumption of a universal energy spectrum of
the electrons. In reality, the spectral index varies from place to
place, and this may induce additional anisotropy. To model this, we
need to include momentum space diffusion. We plan to address
these issues in subsequent studies. 

If the galactic synchrotron emission does have a steep 
angular power spectrum as suggested here, how could we 
reconcile this with the shallower power spectrum reported by the WMAP
team? One intriguing possibility is that there may be 
another type of foreground, which has a frequency dependence similar
to the synchrotron radiation at the relevant bands, and was mistaken as 
synchrotron radiation. Recently, several groups of researchers have
suggested that some of the foreground identified as synchrotron by the
WMAP team may actually be spinning dust \citep{L03,F03,dO04}.  
This has a peak at $10 \GHz - 20 \GHz$, which would not affect the 
21 cm measurements. There might also be other foregrounds, e.g. of 
extragalactic origin, which contributes to the radio
survey. Alternatively, there may be other unknown mechanisms which
produce the small scale anisotropy in the galactic
synchrotron radiation. Further investigations are needed to 
identify what is responsible for producing the small scale anisotropy
power. For the two known physical mechanisms discussed here, the 
anisotropy power at small scales is much smaller than derived from simple
extrapolation.

\acknowledgments
I thank J. L. Han, E. Scannapieco, S. Furlanneto, D. Casadei and
S. P. Oh for suggestions and discussions.
This work is supported by the NSF grant PHY99-07949.

\end{document}